\documentclass[aps,prx,twocolumn,superscriptaddress]{revtex4-1}
\usepackage[latin9]{inputenc}
\usepackage{graphicx}
\usepackage{amsmath}
\usepackage{float}
\usepackage{hyperref}
\usepackage{xcolor}
\usepackage{multirow}

\begin{document}

\title{Plaquette versus ordinary $d$-wave pairing in the $t'$-Hubbard model on a width $4$ cylinder}

\author{Chia-Min Chung}
\affiliation{Niels Bohr International Academy and Center for Quantum Devices,
University of Copenhagen, Lyngbyvej 2, 2100 Copenhagen, Denmark}

\author{Mingpu Qin}
\affiliation{Key Laboratory of Artificial Structures and Quantum Control, School of Physics and Astronomy, Shanghai Jiao Tong University, Shanghai 200240, China}

\author{Shiwei Zhang}
\affiliation{Center for Computational Quantum Physics, Flatiron Institute, New York, NY 10010, USA}
\affiliation{Department of Physics, College of William and Mary, Williamsburg, Virginia 23187, USA} 

\author{Ulrich Schollw\"{o}ck}
\affiliation{Department of Physics and Arnold Sommerfeld Center for Theoretical Physics,
Ludwig-Maximilians-Universit\"{a}t M\"{u}nchen, 80333 Munich, Germany}
\affiliation{Munich Center for Quantum Science and Technology (MCQST), Schellingstrasse 4, 80799 Munich, Germany}

\author{Steven R. White}
\affiliation{Department of Physics and Astronomy, University of California, Irvine, California 92697, USA}

\collaboration{The Simons Collaboration on the Many-Electron Problem}

\begin{abstract}
The Hubbard model and its extensions are important microscopic models for understanding high-$T_c$ superconductivity in cuprates.
In the model with next-nearest-neighbor hopping $t'$ (the $t'$-Hubbard model), pairing is strongly influenced by $t'$.
In particular, a recent study on a width-$4$ cylinder observed quasi-long-rage superconducting order, associated with a negative $t'$, which was taken to imply superconductivity in the two-dimensional (2D) limit.
In this work we study more carefully pairing in the width-$4$  $t'$-Hubbard model.
We show that in this specific system, the pairing symmetry with $t' < 0$ is \emph{not} the ordinary $d$-wave one would expect in the 2D limit.
Instead we observe a so-called plaquette $d$-wave pairing.
The plaquette $d$-wave exists only on a width-$4$ cylinder, and so is not representative of the 2D limit.
We find that a negative $t'$ suppresses the conventional $d$-wave, leading to plaquette pairing.
In contrast, a different $t''$ coupling acting diagonally on the plaquettes suppresses plaquette pairing,
leading to conventional $d$-wave pairing.
\end{abstract}

\maketitle

\section{Introduction}
Understanding the superconductivity in cuprates is one of the greatest challenge in the condensed matter physics~\cite{RevModPhys.66.763}.
The two-dimensional (2D) Hubbard model 
successfully captures most of the important physics, for example the antiferromagnetism at half filling~\cite{ANDERSON1196,PhysRevB.37.3759,RevModPhys.78.17,PhysRevB.31.4403,PhysRevB.94.085140,PhysRevB.94.085103,PhysRevX.5.041041},
and the competing orders under doping~\cite{PhysRevLett.113.046402,PhysRevB.93.035126,Zheng1155,RevModPhys.87.457}.
Among these competing states, the ground state under $1/8$ hole doping in the intermediate coupling regime has been shown to exhibit stripes~\cite{Zheng1155,PhysRevB.97.045138}, where spin density wave and charge density wave (CDW) coexist~\cite{PhysRevLett.91.136403,PhysRevLett.104.116402,PhysRevB.71.075108}.
More precisely, the ground state is a striped state with (CDW) wavelength $\lambda=8$, while the states of $\lambda$ between $5$ and $8$ have extremely close energies.
In experiments $\lambda=8$ stripes were observed recently~\cite{Edkins976}, while the $\lambda=4$ stripes were more widely observed~\cite{Tranquada1995,Tranquada2004,Kohsaka_2007}.

Superconductivity is an even more sensitive property of the model.
The uniform superconducting (SC) state has been shown to exist at higher energy than stripes~\cite{PhysRevLett.113.046402,Zheng1155,PhysRevB.97.045138}.
Whether the superconductivity coexists with stripes is a more subtle question.
Both positive~\cite{PhysRevB.97.075112,PhysRevB.92.195139} and negative~\cite{PhysRevB.95.125125,PhysRevLett.78.4486} results have been reported.
A recent detail study shows that the $d_{x^2-y^2}$-wave SC pairing in the Hubbard model is indeed short ranged at the interaction strengths relevant in the cuprates, and no SC order appears in the thermodynamic limit~\cite{2019Mingpuarxiv:1910.08931}.
This result indicates that the minimal Hubbard model is not sufficient to understand  superconductivity in the cuprates, and one should consider additional terms.

The next-nearest-neighbor hopping $t'$ is an important contribution in the models of cuprates~\cite{ANDERSEN19951573,PhysRevB.98.134501,PhysRevB.99.245155}.
Recent numerical studies showed that the presence of $t'$ drives the wavelength of the stripes from $\lambda=8$ to $4$~\cite{PhysRevB.93.035126,PhysRevB.97.045138,Huang2018}, and enhances superconductivity~\cite{PhysRevB.100.195141,2019arXiv190711728J}.
Specifically, a study of $t'$-Hubbard model on a width-$4$ cylinder observed algebraic decay of SC correlation~\cite{2019Sci...365.1424J}, which may imply true long-range SC order in the 2D limit.

\begin{figure}[h!]
\includegraphics[width=1\columnwidth]{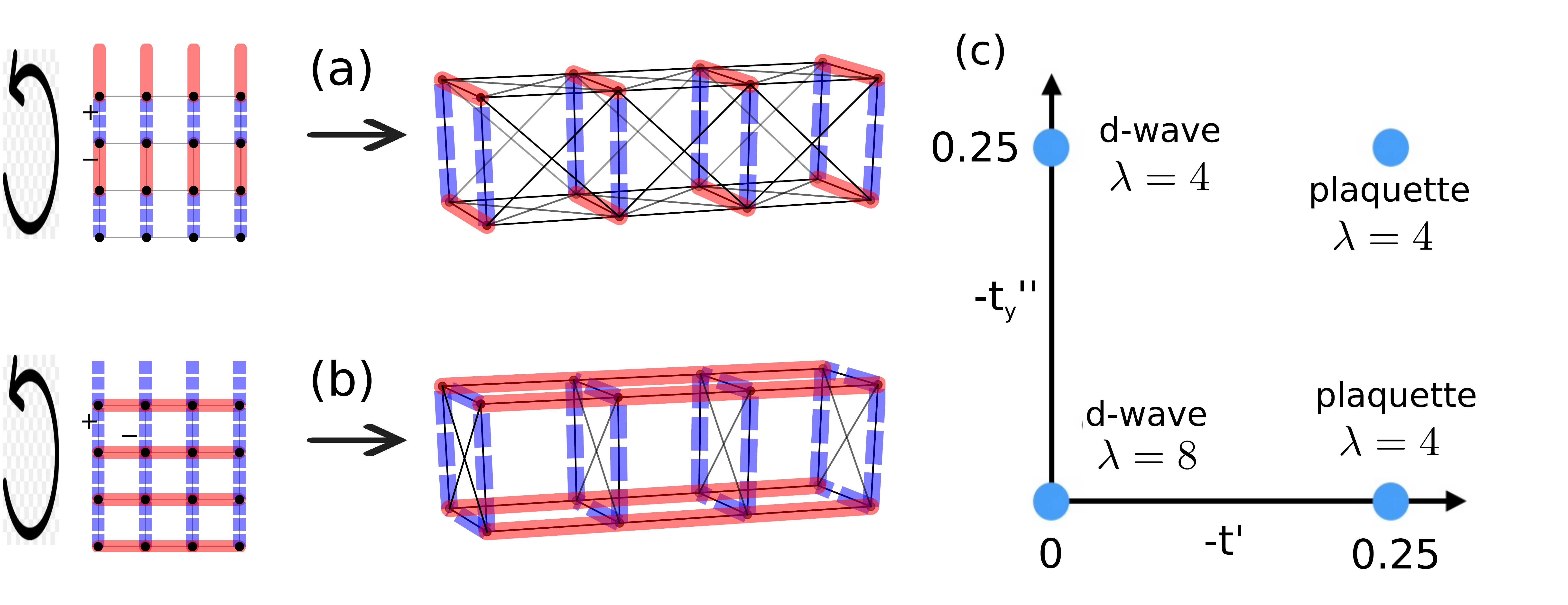}
\caption
{
    (a) Plaquette $d$-wave pairing for $t'\neq 0$ and $t''_y=0$, and (b) ordinary $d$-wave pairing for $t'=0$ and $t''_y\neq 0$ on a width $4$ cylinder.
    The left hand side shows the plane view of a 2D lattice.
    The right hand side shows the view of a stack of plaquettes.
    The red solid (blue dashed) lines represent the positive (negative) sign of the SC pairing.
    (c) A summary of parameters $t'$ and $t''_y$ and their corresponding orders considered here.
}
\label{fig:cartoon}
\end{figure}

In this work we study the $t'$-Hubbard model with negative $t'$ and $1/8$ doping on a width-$4$ cylinder as in Ref.~\cite{2019Sci...365.1424J}.
The ground state is found to show $\lambda=4$ stripes as expected.
However we find that the pairing symmetry is \emph{not} the ordinary $d$-wave that one can extend to the 2D limit.
Instead, due to the fact that the width-$4$ cylinder is identical to a stack of plaquettes, we observe that the ``$d$-wave" pairing arises only on the plaquettes but not along the longitudinal direction, as shown in Fig.~\ref{fig:cartoon}(a).
We call this  {\it plaquette $d$-wave} pairing.

We note that the plaquette $d$-wave pairing has also been observed in the $t'$-$t$-$J$ model on a width-$4$ cylinder~\cite{PhysRevB.95.155116}.
In the Hubbard model without $t'$, the pairing symmetry is the ordinary $d$-wave, and a negative $t'$ drives the system to the plaquette $d$-wave.
In the $t$-$J$ model with $t'=0$, the pairing is the plaquette $d$-wave, and a $t'>0$ drives the system to the ordinary $d$-wave.

We then discuss the effect of $t'$ on the pairing symmetry.
In the view of a stack of plaquettes, the $t'$ hopping exists only on the ``surface" of the cylinder but not on the plaquettes (see Fig.~\ref{fig:cartoon}(b)).
The fact that negative $t'$ drives the $d$-wave from the surface (ordinary $d$-wave) to the plaquettes (plaquette $d$-wave) implies that negative $t'$ \emph{depresses} rather than enhances $d$-wave pairing.
This is consistent with the early studies on the $t'$-$t$-$J$ model~\cite{PhysRevB.60.R753,PhysRevB.79.220504}.
To further examine this idea, we study a system with ``next-nearest" hopping $t''$ only on the plaquettes but not on the surface (see Fig.~\ref{fig:cartoon}(a)).
In other words, we consider a width-$4$ cylinder with $t''_y<0$ and $t'=0$, where $t''_y$ is the third-nearest neighbor in the $y$ direction.
We find that in this system the pairing symmetry becomes the ordinary $d$-wave, consistent with our conjecture.

To further illustrate the uniqueness of the width-4 cylinder, we study a width-6 cylinder for
comparison, as well as a fully open width-4 system (ladder). The plaquette structure is not
present in these systems, and their pairing properties are different. To understand
superconductivity in 2D, systems wider than width-4 must be considered.

\section{System and observables}

The Hamiltonian we consider has the form
\begin{equation}
	\hat{H} =-\sum\limits_{\langle ij\rangle\sigma} t_{ij} \hat{c}_{i\sigma}^{\dagger}\hat{c}_{j\sigma}
     +U\sum\limits _{i}\hat{n}_{i\uparrow}\hat{n}_{i\downarrow}
     -\mu\sum_{i\sigma}\hat{n}_{i\sigma},
	\label{eq:H}
\end{equation}
where $\hat{c}^\dagger_{i\sigma}$ is the Fermionic creation operator, $\sigma=\{\uparrow,\downarrow\}$ denotes spin, and $\hat{n}_{i\sigma}=\hat{c}_{i\sigma}^\dagger\hat{c}_{i\sigma}$ is the particle-number operator.
In this work we consider mainly $t_{ij}=t$ and $t'$ for the nearest and next-nearest neighbors respectively,  and $t_{ij}=0$ for other terms, except in one special toy case, where we set 
$t_{ij}=t''_y\neq 0$ for the third-nearest neighbors in the $y$ direction, while setting $t'=0$.
The chemical potential term is included only when an additional pairing field is applied to the system (explained below) and thus the particle number is not conserved.
Throughout this paper, $t=1$ is the energy unit.
We consider cylindrical boundary conditions, where the system is open in the x- and periodic in the y-direction,
as illustrated in Fig. 1.
In Sec. IV fully open boundary conditions are also considered for comparison.

We are mostly interested in the SC properties of the ground state.
We define the (singlet) pairing operator on two nearest-neighbor sites $i$ and $j$ (or a bond),
\begin{equation}
	\hat{\Delta}_{ij}\equiv\frac{(\hat{c}_{i\uparrow}\hat{c}_{j\downarrow}-\hat{c}_{i\downarrow}\hat{c}_{j\uparrow})}{\sqrt{2}}
	\label{eq:Delta}
\end{equation}
The coherence of the pairs is measured by the pair-pair correlation between two bonds
\begin{equation}
	P(i',j';i,j)=\langle \hat{\Delta}_{i'j'}^{\dagger} \hat{\Delta}_{ij} \rangle.
	\label{eq:ppcorr}
\end{equation}
The strongest order the system can have on the cylinder is quasi-long-range order,
where the correlation decays algebraically.
Alternatively, if the system has no SC order, the correlation decays exponentially.
Another useful probe to detect SC is to apply a weak pairing field $\sum_{ij} h_p^{ij}\hat{\Delta}_{ij}$ to the system, and measure the induced SC order $\langle \Delta_{ij}\rangle$.
Without a pairing field the Hamiltonian conserves particle number, while in the presence of a
pairing field it does not.

The density-matrix renormalization group (DMRG) method is a variational method that approximates the ground state due to its low-entanglement~\cite{PhysRevB.48.10345,PhysRevLett.69.2863,SCHOLLWOCK201196}.
These quasi-1D systems can be accurately simulated by DMRG.
We use two different setups in DMRG for two types of systems.
For the systems without pairing field, in studying the correlations, we use DMRG with conserved particle number and $SU(2)$ symmetry, together with a single-site update~\cite{hubig17:_symmet_protec_tensor_networ,PhysRevB.91.155115}.
For the system with pairing field, in studying the SC order, we use DMRG without particle-number conservation, and with a two-site update.

The constrained path auxiliary-field quantum Monte Carlo (AFQMC)~\cite{PhysRevB.55.7464,PhysRevB.94.235119} has an approximation completely independent of entanglement, providing a good complementary crosscheck with DMRG.
For pairing measurements with AFQMC, we use  the Hellmann-Feynman formula to convert an energy measurement to
a measurement of the pairing order parameter, significantly increasing the accuracy~\cite{2019Mingpuarxiv:1910.08931}.

\section{Pair-pair correlation}

\begin{figure}[h!]
\includegraphics[width=1\columnwidth]{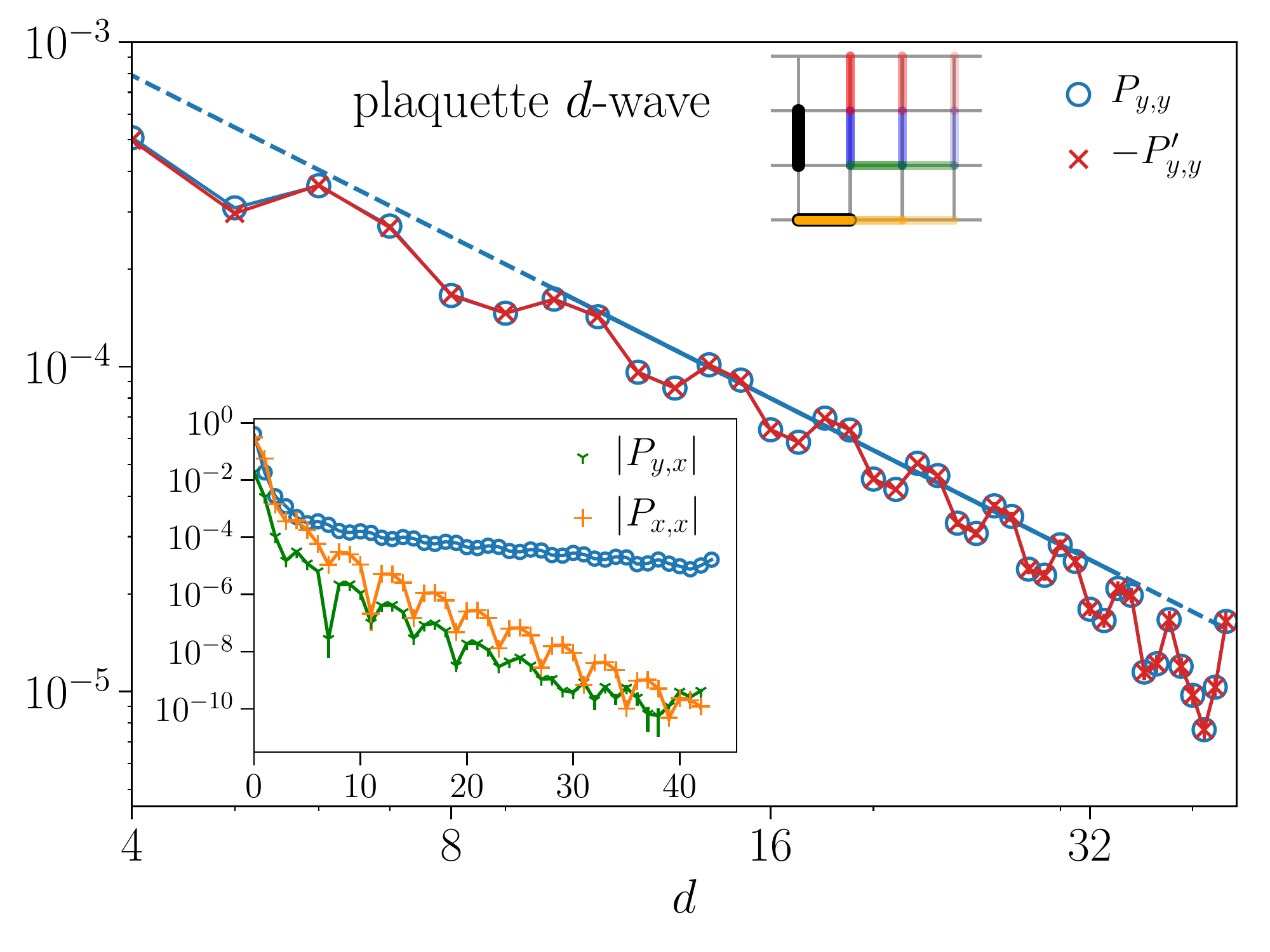}
\caption
{
    Pair-pair correlations $P=\langle\Delta_{i'j'}^\dagger \Delta_{ij}\rangle$ as functions of distance $d$.
    The system is a $48\times 4$ cylinder with $t'=-0.25$.
    Different types of correlations are shown:
    $P_{y,y}$ (blue), $P'_{y,y}$ (red), and $P_{y,x}$ (green) are the correlations of the blue, red, and green bonds with respect to the black bond.
    $P_{x,x}$ is the correlation of the thin yellow bonds with respect to the thick yellow bond.
    The main panel shows $P_{y,y}$ and $-P'_{y,y}$ in log-log scale.
    The blue solid line is the algebraic fit to the peak values of $P_{y,y}$ for $d$ from $10$ to $30$, and the dashed line is an extension to guide the eye.
    The inset shows $|P_{y,x}|$ and $|P_{x,x}|$ compared with $P_{y,y}$ in semi-log scale.
}
\label{fig:pcorr_tp}
\end{figure}
We first discuss the pair-pair correlation in the ground state of the $t'$-Hubbard model for $t'=-0.25$ ($t''_y=0$) on a width-$4$ cylinder. We consider $1/8$ hole doping throughout if not stated otherwise.
To characterize different pairing symmetries, we measure four types of the pair-pair correlation in Eq.~(\ref{eq:ppcorr}),  which we will denote by $P_{\alpha,\beta}(d)$, with $\alpha$ and $\beta$ ($=x$ or $y$) giving the character of the bonds $\langle ij\rangle$ and $\langle i'j'\rangle$, and $d$ being the distance between $i$ and $i'$.
(We use the convention that, for each bond $\langle ij\rangle$, $j$ is either ``above'' or  ``to the right" of $i$, for a vertical or horizontal bond.)
Thus $P_{y,y}(d)$, $P_{x,x}(d)$, and $P_{y,x}(d)$ denote the correlations between two vertical, two horizontal, and one vertical (at $i$) and one horizontal (at $i'$) bonds, respectively, with the bonds located at $i=(5,y_0)$ and $i'=(5+d,y_0)$; 
$P'_{y,y}(d)$ will denote the correlation between two vertical bonds at $i=(5,y_0)$ and $i'=(5+d,y_0+1)$, i.e.\ shifted by one site vertically.
See the diagram in Fig.~\ref{fig:pcorr_tp} for a sketch.

In Fig.~\ref{fig:pcorr_tp} we show these correlations as functions of the distance $d$ to the reference bonds.
One can see that $P_{y,y}$ and $-P'_{y,y}$, the two kinds of vertical-vertical correlations with the opposite signs, are perfectly symmetric.
This indicates the $d_{x^2-y^2}$-wave pairing on the plaquettes (see Fig.~\ref{fig:cartoon}(a)).
The inset shows comparisons of $|P_{y,x}|$ and $|P_{x,x}|$ to $P_{y,y}$.
It can be seen that both $|P_{y,x}|$ and $|P_{x,x}|$ decay exponentially and much faster than $P_{y,y}$, showing that the pairing is only on the plaquettes but not along the longitudinal direction.
In fact $P_{y,x}$ and $P_{x,x}$ oscillate up and down around zero (hence we plot their absolute values).
Together these correlations show that the pairing symmetry is plaquette $d$-wave.
We employ an algebraic fit $P_{y,y}\propto d^{-K_{\mathrm{SC}}}$ for the peak values of $P_{y,y}$ for $d$ from $10$ to $30$, and obtain $K_{\mathrm{SC}}=1.62(5)$ which is compatible with what was obtained in Ref.~\cite{2019Sci...365.1424J}.

We see that only measuring the correlation of two vertical bonds at the same $y$ position is not able to distinguish the plaquette $d$-wave and the ordinary $d$-wave, which show fundamentally different properties.
It is necessary to measure different kinds of correlations to understand the pairing symmetry.

\begin{figure}[h!]
\includegraphics[width=0.9\columnwidth]{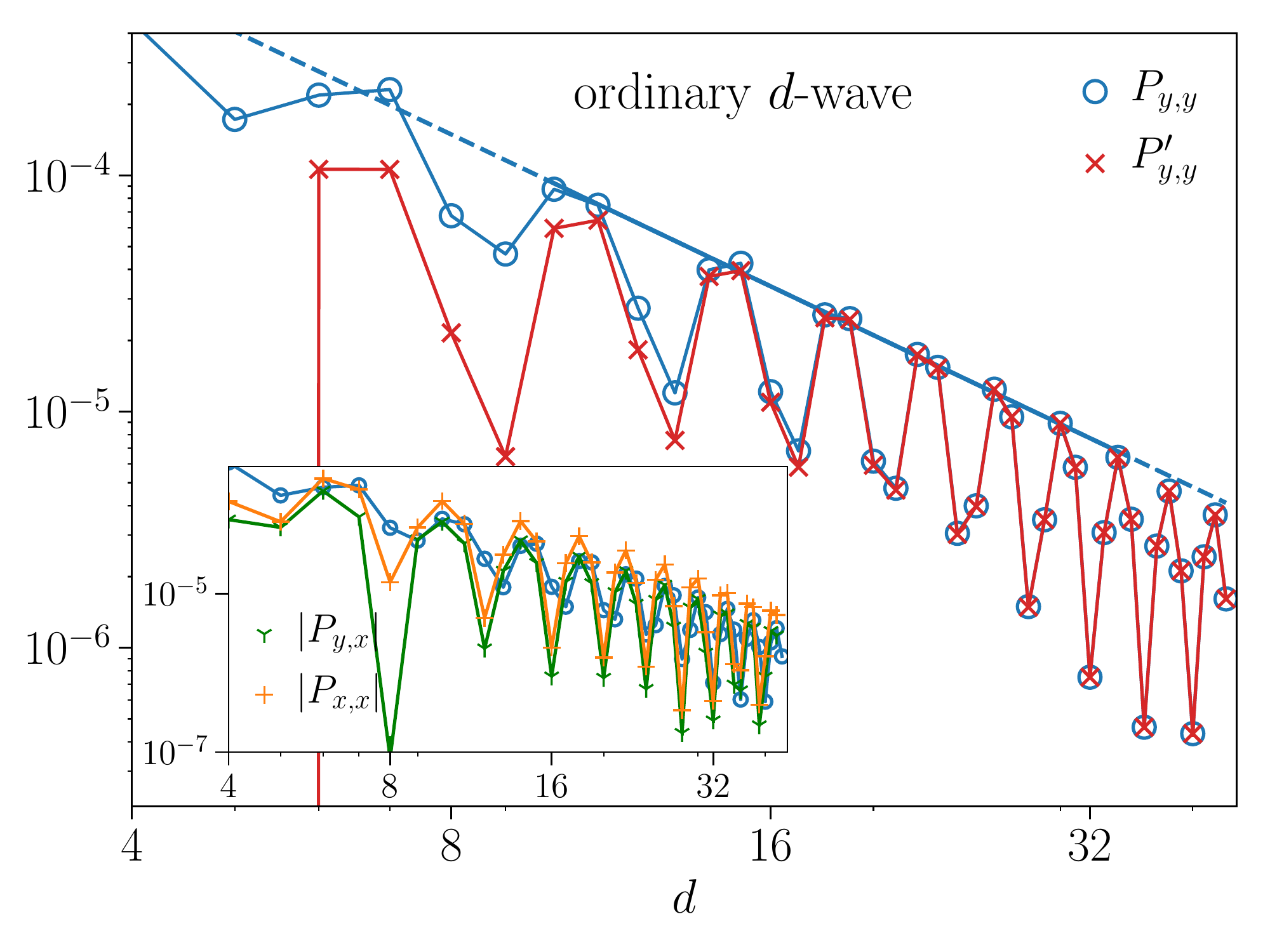}
\caption
{
    Pair-pair correlations $P=\langle\Delta_{i'j'}^\dagger \Delta_{ij}\rangle$ as functions of distance $d$ in log-log scale.
    The system is a $48\times 4$ cylinder with $t''_y=-0.25$ and $t'=0$.
    Different types of correlations are shown, as explained in the caption of Fig.~\ref{fig:pcorr_tp}.
    The blue solid line is the algebraic fit to the peak values of $P_{y,y}$ for $d$ from $10$ to $30$, and the dashed line is just the extension to guild to eyes.
}
\label{fig:pcorr_plaque}
\end{figure}
\paragraph*{Next-neighbor hopping on plaquettes.}
Recall that $t'$ only includes next-nearest hopping on the surface but not on the plaquettes of the cylinder.
The fact that $t'<0$ drives the $d$-wave pairing from the surface to the plaquettes implies that $t'<0$ tends to \emph{suppress} the $d$-wave pairing, therefore pushes the $d$-wave from the surface to the plaquettes.
To further examine this idea, we consider a system with the ``next nearest" hopping \emph{only} on the plaquette.
More precisely, we consider $t''_y=-0.25$ and $t'=0$; see Fig.~\ref{fig:cartoon}(b).
The ground state still has stripes with $\lambda=4$; however the pairing symmetry changes.
The different types of pair-pair correlations are shown in Fig.~\ref{fig:pcorr_plaque}.
In this system the pairing symmetry is back to the ordinary $d$-wave:
$P_{y,y}$ and $P'_{y,y}$ are symmetric with the \emph{same} sign,
and $|P_{y,x}|$ and $|P_{x,x}|$ are compatible with $|P_{y,y}|$.
The algebraic fit $P_{y,y}\propto d^{-K_{\mathrm{SC}}}$ for the peak values $P_{y,y}$ gives $K_{\mathrm{SC}}=2.13(5)$.
The result is consistent with our conjecture that the next-nearest hopping $t'<0$ \emph{suppresses} $d$-wave pairing, and therefore the $t''_y$ on the plaquettes stabilize the ordinary $d$-wave on the surface.

In Fig.~\ref{fig:cartoon}(c) we summarize the parameters $t'$ and $t''_y$ and their corresponding pairing symmetries that we consider in this work. Note that the results of $t=t'=0$ and $-0.25$ are not shown in the paper but have been studied.

\section{Other Cluster Geometries}
In this section we illustrate that the plaquette $d$-wave pairing can only exist on a width-$4$ cyinder due to its special geometry.
It does not exist in a system with open boundaries in both directions, and cannot be extended to a width-$6$ cylinder.

\begin{figure}[h!]
\includegraphics[width=1\columnwidth]{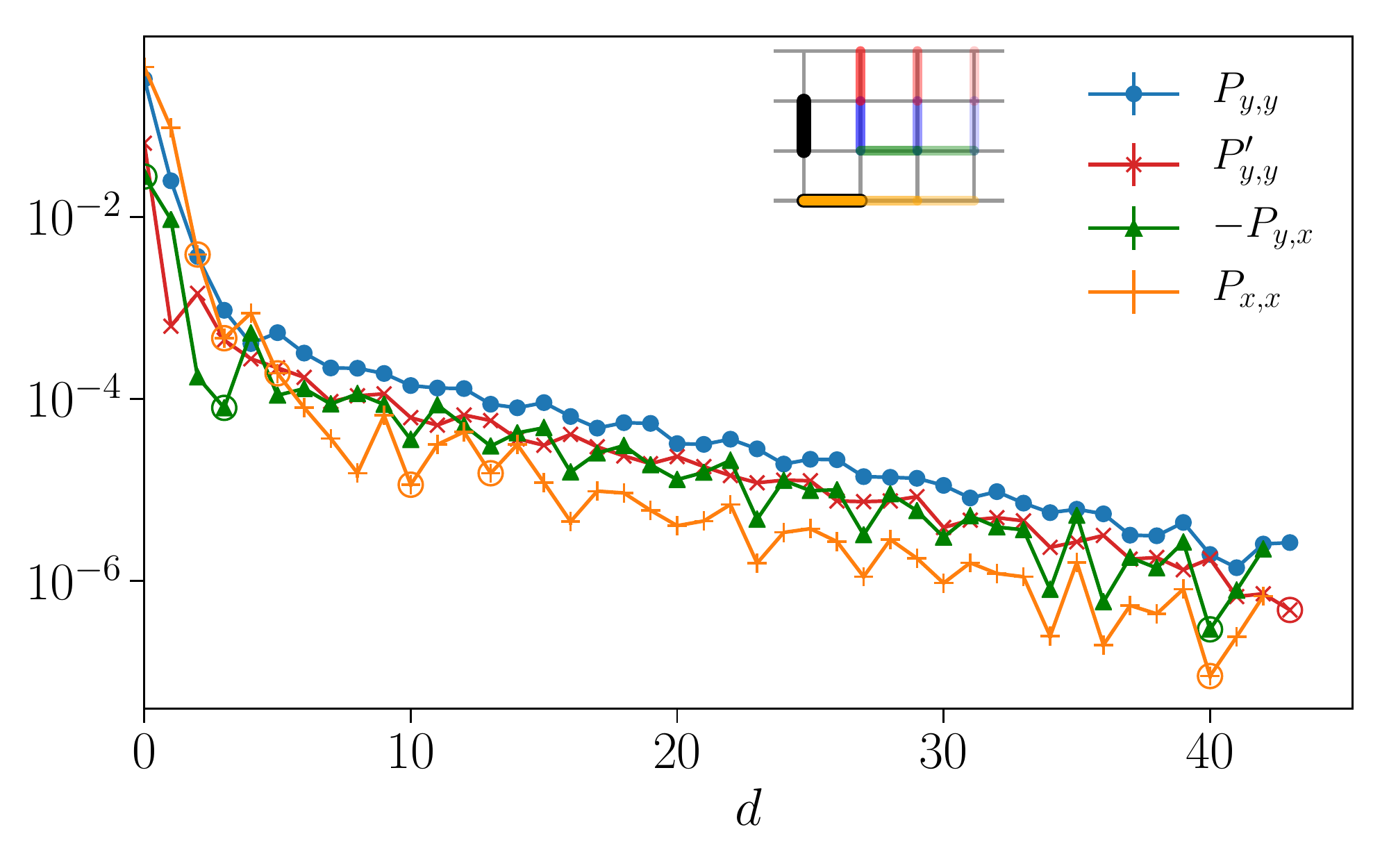}
\caption
{
    Pair-pair correlations $P=\langle\Delta_{i'j'}^\dagger \Delta_{ij}\rangle$ as functions of distance $d$ for $t'=-0.25$ for a $48\times 4$ system with open boundaries.
    Different types of correlations are shown, as explained in the caption of Fig.~\ref{fig:pcorr_tp}.
    Note that the signs of the correlations are different from those in Fig.~\ref{fig:pcorr_tp}.
}
\label{fig:pcorr_obc}
\end{figure}
\paragraph*{Fully open boundary conditions.}
The plaquette structure is specific to the cylindrical boundary condition on a width-4 cylinder.
With open boundaries in both directions, the vertical bonds along the circumferential direction do not form a plaquette, and therefore the system is expected to \emph{not} have plaquette $d$-wave pairing.
Here we consider a $48\times 4$ system with open boundaries for $t'=-0.25$.
The pair-pair correlations are shown in Fig.~\ref{fig:pcorr_obc}.
One can see that the pairing symmetry is the ordinary $d$-wave:
$P_{y,y}$ has the same (opposite) sign with $P'_{y,y}$ ($P_{y,x}$).
The correlations are not perfectly symmetric due to the open boundaries in the $y$ direction.

\begin{figure}[h!]
\includegraphics[width=0.9\columnwidth]{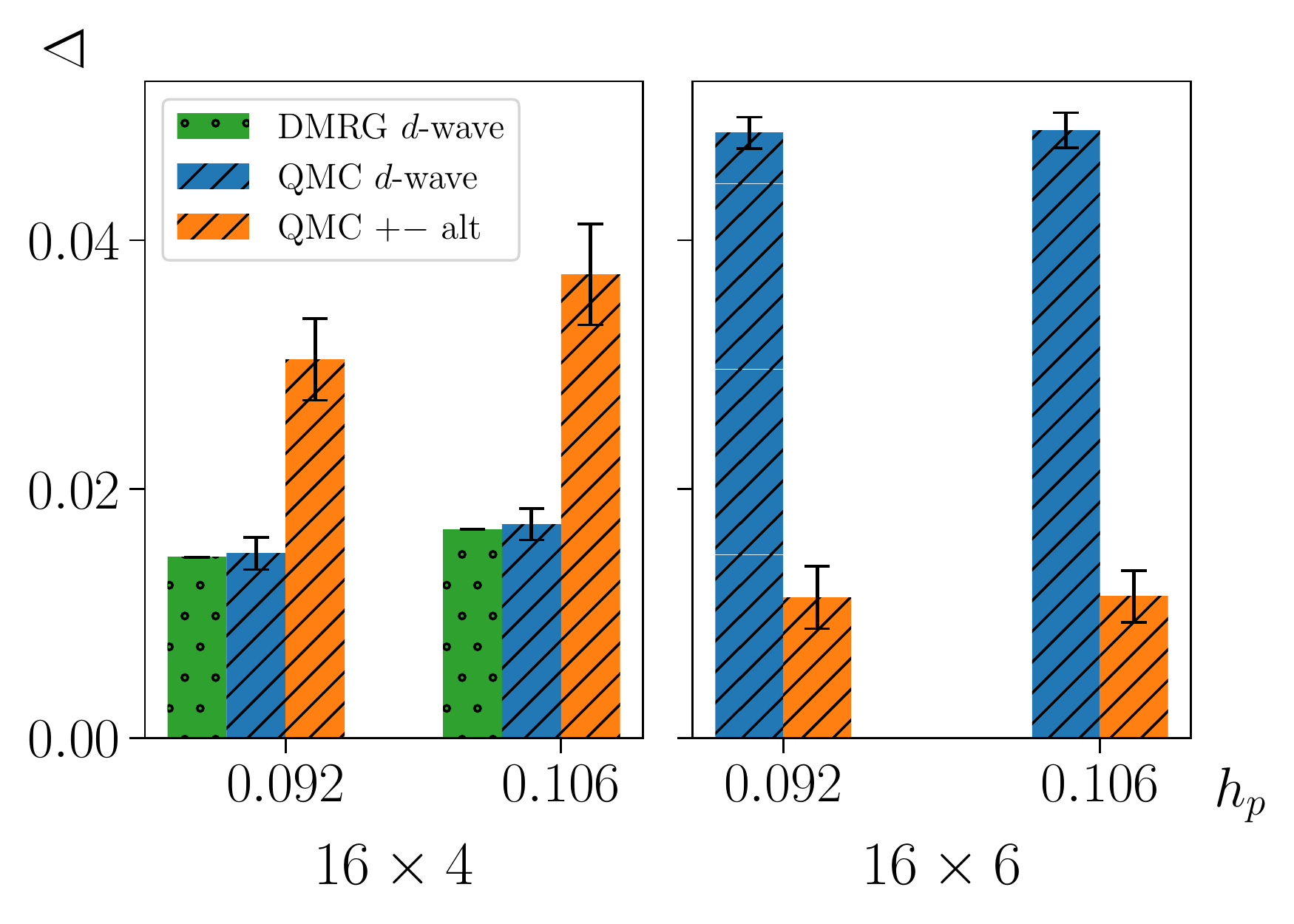}
\caption
{
    SC orders induced by two different types of pairing fields with magnitude $h_p$ for $16\times 4$ and $16\times 6$ cylinders.
    The blue (orange) bars are for the ordinary $d$-wave ($+-$ alternating) pairing field from QMC.
    The green bars are for the ordinary $d$-wave pairing field from DMRG.
}
\label{fig:delta_qmc}
\end{figure}

\paragraph*{A $16\times 6$ cylinder.}
Here we discuss the pairing symmetry in a $16\times 6$ cylinder for $t'=-0.32$,
using AFQMC to study the SC order induced by pairing field.
To see if the simple extension of the plaquette $d$-wave can be applied to a wider cylinder, 
we consider two types of pairing fields $\sum_{\langle ij\rangle} h_p^{ij}\hat{\Delta}_{ij}$ which will induce two types of pairing symmetries:
1) the ordinary $d$-wave pairing field applied on all bonds, with $h_p^{ij}=h_p$ and $-h_p$ for the vertical and horizontal bonds, respectively;
2) the ``$+-$" alternating pairing field applied only on the vertical bonds, with $h_p^{ij}=h_p$ for the odd and $-h_p$ for even vertical bonds.
The alternating pattern in (2) leads to the plaquette $d$-wave on a width-4 cylinder,
and is a ``natural" (but not unique) extension of the plaquette pattern to a width-$6$ cylinder.
Fig.~\ref{fig:delta_qmc} shows the SC orders induced by the two types of pairing field for $16\times 4$ and $16\times 6$ cylinders.
(We verify the accuracy of AFQMC by comparing to the DMRG data for $16\times 4$ cylinder with ordinary $d$-wave pairing field,
where excellent agreement is seen between the two methods.)
It can be seen that, for a $16\times 4$ cylinder, the plaquette $d$-wave SC order is stronger than the ordinary $d$-wave,
while for a $16\times 6$ cylinder, the ordinary $d$-wave is stronger.
This result shows that the plaquette $d$-wave pairing is special for a width-$4$ cylinder.

\section{Conclusions}
In this work we have studied the SC pairing symmetry in the ground state of $t'$-Hubbard model for $t'<0$.
We have shown that the pairing symmetry in a width-$4$ cylinder is the plaquette $d$-wave rather than the ordinary $d$-wave, and therefore {\em this system is not representative of the 2D limit.}
In contrast, one sees ordinary $d$-wave pairing,  on a width-$6$ cylinder.
However, even width 6 is still limited by the quantized fillings of the short stripes that circle the cylinder, which are observed to always contain an even number of holes. Thus a half-filled stripe is not seen in width 6, which would require 3 holes.  
Ideally, to establish the nature of the superconducting correlations in 2D,
a systematic approach to the thermodynamic limit will be very important.
We also discuss the role of $t'$ for $d$-wave pairing.
We find that on the width 4 cylinder $t'<0$ suppresses rather than enhances the $d$-wave pairing.
This result is \emph{inconsistent} with a recent infinite projected entangled pairs study~\cite{PhysRevB.100.195141} focusing on the 2D limit. 
It may be that the role of $t'$ is subtle, both weakening stripes (and thus favoring pairing) but also weakening the pair in a more direct fashion.

\section{Acknowledgements}

U.S. acknowledge support by the Deutsche Forschungsgemeinschaft (DFG, German Research Foundation) under Germany's Excellence Strategy -- 426 EXC-2111 -- 390814868. SRW acknowleges the support of the NSF under NSF DMR-1812558. Part of the work was carried out at the Flatiron Institute. The Flatiron Institute is a division of the Simons Foundation.

\bibliography{hubbardtpx4}

\appendix

\section{$t'$ dependence}
\begin{figure}[h!]
\includegraphics[width=1\columnwidth]{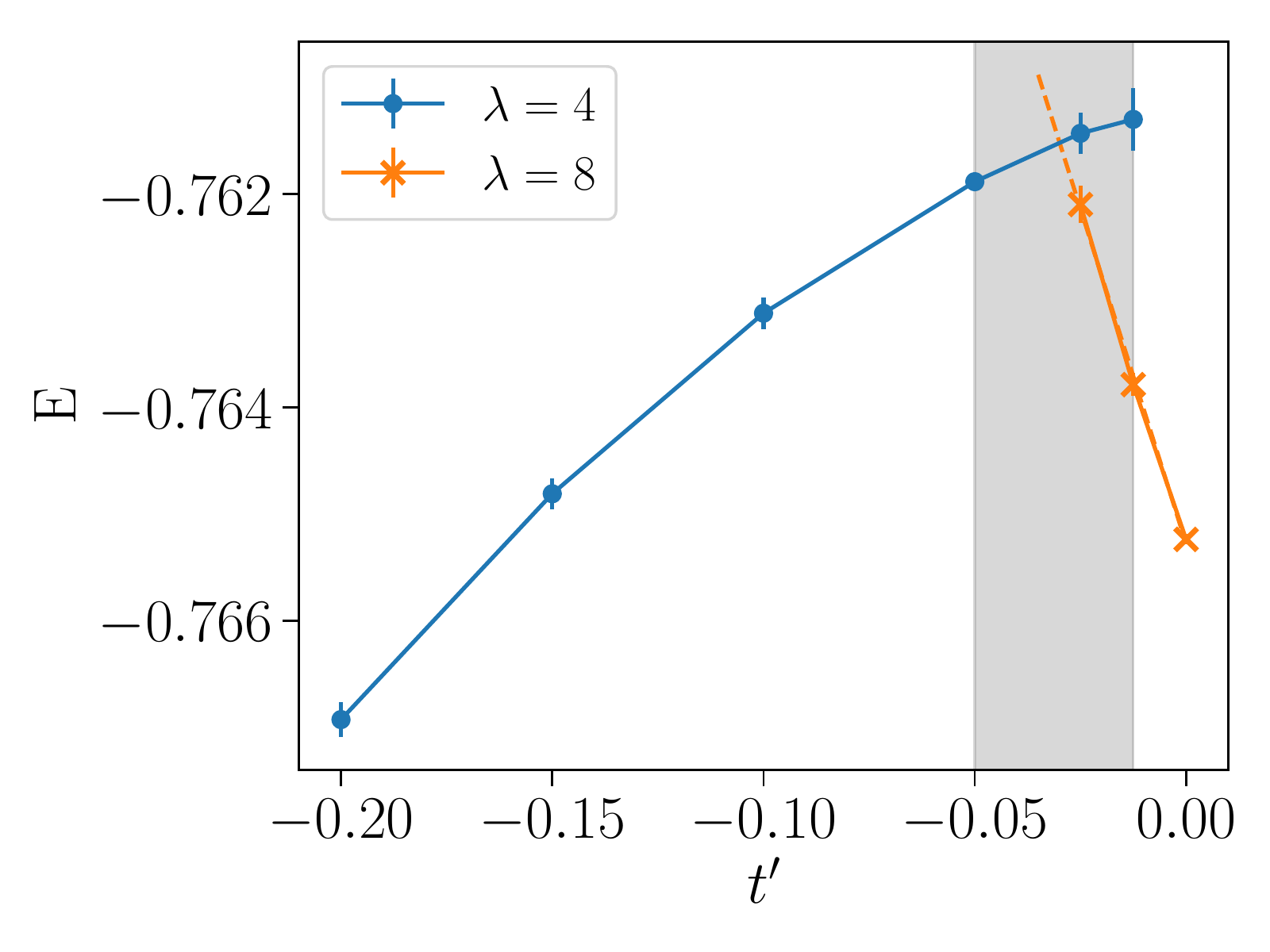}
\caption
{
    The energies of the filled ($\lambda=8$) and the half-filled ($\lambda=4$) stripes on a width-$4$ and infinite-length cylinder for different $t'$.
    The gray area indicates the region where the ground state is the mixture of these two striped states.
    The orange dashed line is a linear extrapolation of the orange crosses.
}
\label{fig:tp_dep}
\end{figure}
Here we study the transition from filled ($\lambda=8$) stripes to half-filled ($\lambda=4$) stripes induced by $t'$ on a width-$4$ and infinitely long cylinder.
A more complete phase diagram for $t'$ and $U$ has been addressed in Ref.~\cite{2019arXiv190711728J}.
Note that the pairing symmetries are different in these two striped states:
The filled stripes show the ordinary $d$-wave pairing and the half-filled stripes have plaquette $d$-wave pairing.
In Fig.~\ref{fig:tp_dep} we show the energies of filled and half-filled stripes for different $t'$.
The energies have been extrapolated to infinite system length.
For small (large) negative $t'$, the ground state shows filled (half-filled) stripes.
The gray area shows the region where a mixture of filled and half-filled stripes is seen in the ground states.

\section{U dependence}
\begin{figure}[h!]
\includegraphics[width=1\columnwidth]{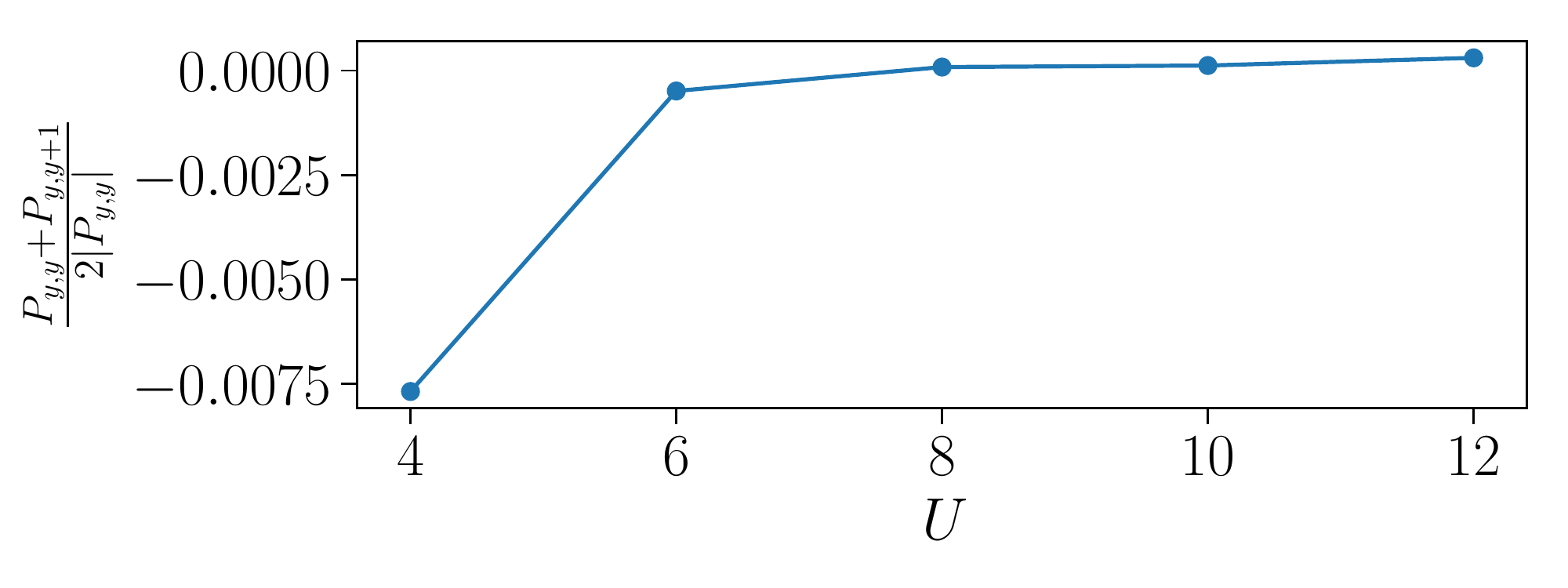}
\caption
{
    Sum of two types of vertical correlations, $\frac{P_{y,y}+P'_{y,y}}{2|P_{y,y}|}$, at distance $d=L/2$.
    The system is a $24\times 4$ cylinder with $t'=-0.25$.
    If the system has perfect plaquette (surface) $d$-wave pairing, this quantity should equal to zero (one).
}
\label{fig:U_dep}
\end{figure}
We examine the $U$-dependence of the plaquette $d$-wave pairing for $t'=-0.25$.
Since the pairing symmetry is not sensitive to the length of the system, we focus on a $24\times 4$ cylinder.
To characterize the different pairing symmetries, we compute the normalized sum of the two types of vertical correlations, $\frac{P_{y,y}+P'_{y,y}}{2|P_{y,y}|}$, at distance $d=L/2$.
For the perfect plaquette (surface) $d$-wave pairing, $P_{y,y}$ would have opposite (the same) sign with $P'_{y,y}$, and this normalized sum will equal to zero (one).
As shown in Fig.~\ref{fig:U_dep}, the plaquette $d$-wave pairing is very stable in all $U$ we study here from $U=4$ to $U=12$.

\end{document}